\documentclass[USenglish,twocolumn]{article}
\usepackage[utf8]{inputenc}
\usepackage[big,online]{dgruyter}
\usepackage[sort&compress,square,numbers]{natbib}
\usepackage{times}
\usepackage{xcolor}%

\begin{document}

  \DOI{10.1515/}
  \openaccess
  \pagenumbering{gobble}

\title{Reduction of Electromagnetic Interference in ultra-low noise Bimodal MEG \& EEG}
\runningtitle{Reduction of EMI in ultra-low noise bimodal MEG \& EEG}

\author*[1]{Jim Barnes}
\author[3]{Lukasz Radzinski}
\author[2]{Soudabeh Arsalani}
\author[3]{Gunnar Waterstraat}
\author[3]{Gabriel Curio}
\author[4]{Jens Haueisen}
\author[2]{Rainer Körber}
\runningauthor{J. Barnes et al.}

\affil[1]{\protect\raggedright 
  Physikalisch-Technische Bundesanstalt, Abbestraße 2-12, 10587 Berlin, Germany, e-mail: jim.barnes@ptb.de}
\affil[2]{\protect\raggedright 
  Physikalisch-Technische Bundesanstalt, Abbestraße 2-12, 10587 Berlin, Germany}
\affil[3]{\protect\raggedright 
  Charit\'e-Universitätsmedizin Berlin, 12203 Berlin, Germany}
\affil[4]{\protect\raggedright
  Technische-Universität Ilmenau, 98693 Ilmenau, Germany}	

\abstract {Single-channel SQUID system technology, operating at a noise level of 100's of aT/${\sqrt{\textrm{Hz}}}$, enables the non-invasive detection of synchronized spiking activity at the single-trial level via magnetoencephalography (MEG). However, when combined with simultaneous electroencephalography (EEG) recordings, the noise performance of the ultra-sensitive MEG system can be greatly diminished. This issue negates some of the complementary qualities of these two recording methods. In addition, typical electrical components required for electrical stimulation of peripheral nerves, a common method for evoking specific brain responses, are also observed to have a detrimental influence on ultra-low MEG noise performance. These effects are caused by electromagnetic interference (EMI) and typically preclude single-trial detection. This work outlines, how careful design allows a significant reduction of the impact of EMI when these different electronic systems are operated concurrently. This optimization enabled the simultaneous single-trial detection of synchronized spiking activity using these two highly sensitive recording modalities.}


\maketitle

\section{Introduction} 

The combination of magneto- and electroencephalography (MEG/EEG) for simultaneous recording is a common paradigm in the field of neurology~\cite{Hari2017, Haueisen2002, Aydin2014}. The low electrical conductivity of the skull and scalp strongly influences the electrical potential at the head surface and limits accurate source localization in EEG. On the other hand, the magnetic permeability of these body parts is nearly that of free space and consequently the magnetic field is not distorted in MEG \cite{Vorwerk2014, Singh2014}. This means MEG is less dependent on volume conductor modeling than EEG. However, MEG, unlike EEG, is unable to detect radial currents \cite{Haemaelaeinen1993}. Therefore, their combination can be used to obtain more accurate source localization \cite{Sharon2007}. A significant advantage of both methods is their high temporal resolution in the millisecond range.

Most commercially available low-noise SQUID-MEG devices operate at noise levels of around $\rm 2$ fT/${\sqrt{\textrm{Hz}}}$, and any noise increase with simultaneous EEG recording is almost negligible. While these devices have been used successfully for the investigation of post synaptic potentials with frequencies of up to some hundred Hz, they are unable to detect high frequency oscillations (HFO) > 500~Hz without averaging. Recent advances in ultra-sensitive, single-channel SQUID system technology achieved white noise levels below $\rm 300$ aT/${\sqrt{\textrm{Hz}}}$~\cite{Storm2017, Storm2019} and enabled the single-trial detection of high frequency somatosensory evoked responses (hfSERs) following electrostimulation of the median nerve~\cite{Waterstraat2021}. Of particular interest is the so-called $\sigma$-range of $\rm 450-850 Hz$ as they represent a macroscopic marker of synchronized cortical spiking activity~\cite{Baker2003} and its single-trial detection allows the non-invasive investigation of neuronal processing. Ultimately, a multi-channel MEG system combined with multi-channel EEG would allow for spatially resolved measurements of this spiking activity, further reducing the gap between invasive and non-invasive measurements. 

In this work, we show that the combination of ultra-sensitive MEG and EEG leads to significant technical challenges. Careful optimization of the entire MEG/EEG hardware is required to maintain an MEG noise performance below $\rm 500$ aT/${\sqrt{\textrm{Hz}}}$ in the $\sigma$-range. The minimization of these destructive noise effects allowed us to perform simultaneous bimodal MEG/EEG recordings of hfSERs at the single-trial level.

\section{Methodology}
In our setup for simultaneous ultra-sensitive MEG/EEG recordings, an 8-channel EEG is mounted onto the region close to the somatosensory cortex with the single-channel MEG sensor positioned above. The electrode leads run from the subject's head under the MEG sensor to a breakout box inside a moderately magnetically shielded room (MSR)~\cite{Voigt2013}. A (ribbon) cable is used to attach this breakout box to the low-noise EEG amplifier~\cite{Scheer2011} with in-built data acquisition outside the MSR. An optical cable feeds into an interface which connects the control laptop via an USB cable. The entire hardware is located inside an RF shielded room.

\begin{figure}
\centering
	\includegraphics[width=0.8\columnwidth]{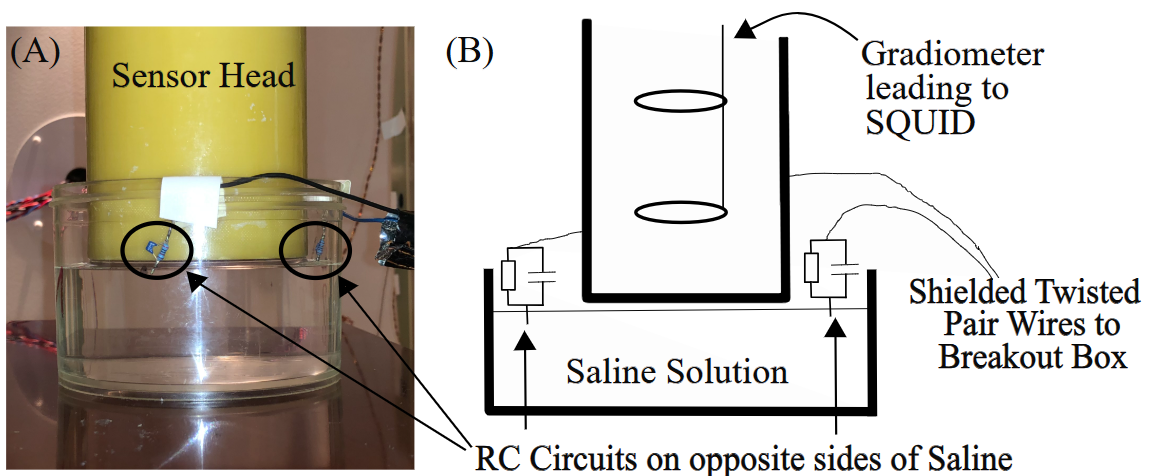}
	\caption{Basic 2-channel EEG phantom used for optimization of the combined low noise EEG \& MEG setup. (A): Picture showing the low noise SQUID sensor head above the EEG phantom. (B): Circuit Schematic  of the EEG phantom.}
	\label{fig:Figure1}
\end{figure}

For optimization, we reproducibly recreated an EEG recording setup by deploying a basic EEG phantom, see Figure~\ref{fig:Figure1}. A cylindrical container was filled with a saline solution of 0.17\% NaCl which mimics the conductivity of the brain $\rm 0.29 \pm 0.005 S/m$~\cite{Hunold2020}. For low-noise EEG, an electrode-scalp impedance of $\rm 1k\Omega$ can be achieved~\cite{Waterstraat2015a}, while the capacitance between the electrode gel and the scalp is roughly $\rm 50nF$~\cite{Heikenfeld2018}. Hence, a $\rm 47 nF$ capacitor was connected in parallel to a $\rm 1 k\Omega$ resistor with one end submerged into the solution. This was done twice to recreate a 2-channel EEG, one signal and one reference channel, enabling the reproducible generation of EEG noise effects without requiring a subject.

A thorough examination of the stimulation system, together with its trigger recording, was also required to identify sources of EMI from the stimulator hardware to the ultra-low noise SQUID system which can easily occur. Of particular importance is the avoidance of radio frequency interference (RFI) which has been shown to increase the SQUID noise~\cite{Koch1994, Ishikawa1993}. Through multiple concurrent recordings of MEG/EEG with the stimulation electronics operated, sources of interference were determined and mostly eliminated by adhering to the following guidelines:

\begin{itemize}
\item Provide proper shielding to cables entering the MSR with good electrical connection to the system ground. 
\item Keep electronic devices and electronic systems as far apart as practicable to prevent electronic coupling (EC).
\item Avoid where possible parallel cables to reduce EC.
\end{itemize}

With the optimized setup, we then performed concurrent low-noise 8-channel EEG and single-channel MEG for subjects during resting state (subject in position under dewar, no stimulation occurring) and following electrostimulation of the median nerve to detect hfSERs at the single-trial level for both modalities.

\begin{figure}
\centering
	\includegraphics[width=0.7\columnwidth]{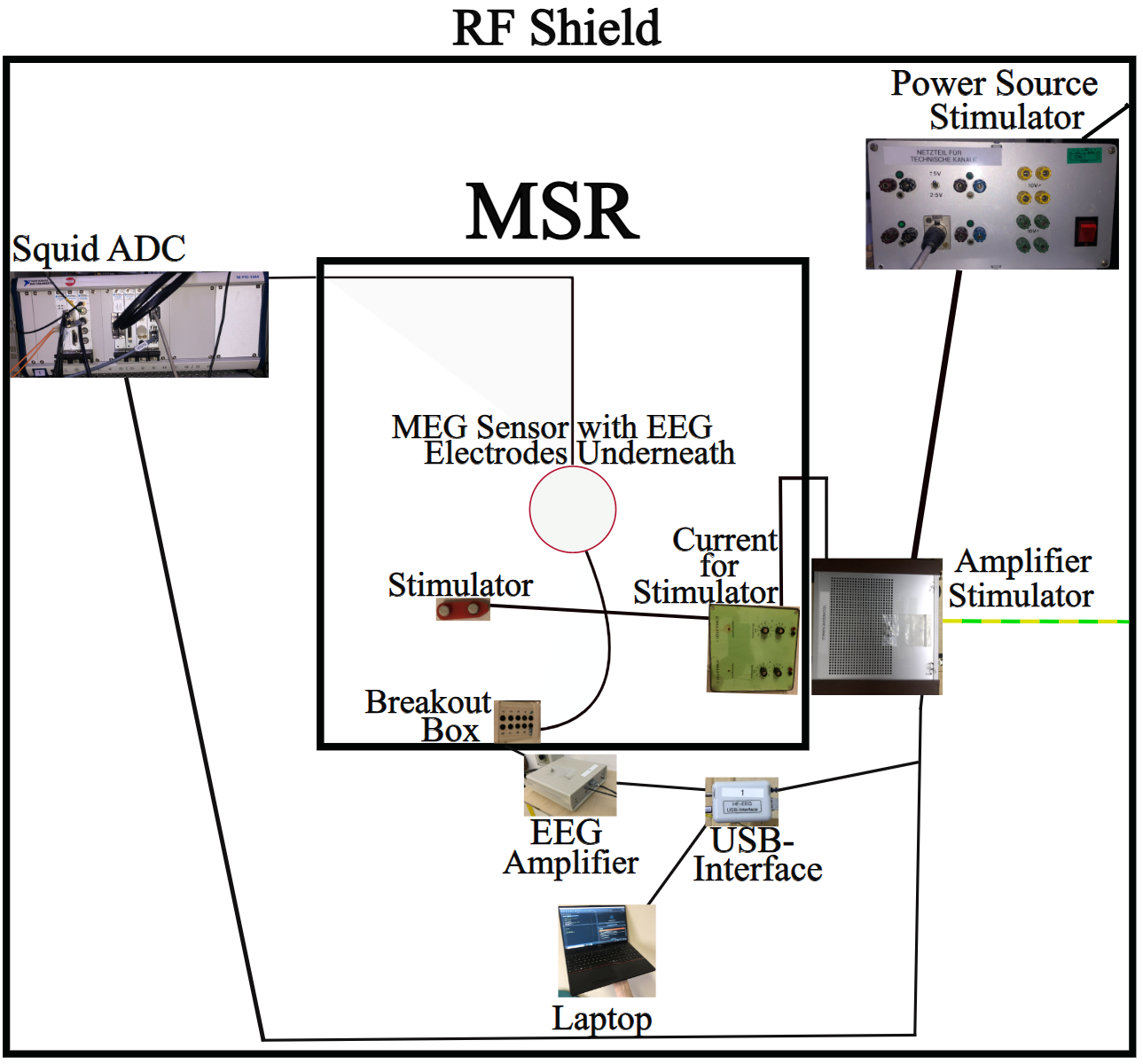}
	\caption{Schematic of the optimized MEG/EEG setup operated inside an MSR and RF shielded room. Physical separation of SQUID and stimulation hardware results in minimal coupling and is required for ultra-low noise performance. }
	\label{fig:Figure2}
\end{figure}

\section{Results}
\subsection{Minimization of EMI}
We first describe the measures which led to significant improvements in MEG noise performance. This is not a complete description but gives the major contributions. A schematic of the noise optimized combined MEG/EEG setup is shown in Figure~\ref{fig:Figure2}. 

The connecting cable into the MSR between the EEG amplifier and the break out box is critical and requires good electrical shielding and an RF-tight feed-through. This is the case as some of the EEG electrodes are positioned directly under the MEG system and can therefore easily couple RFI to the SQUID system. In addition, guiding the EEG electrodes cables away from the sensor head (preferably to one side) helps to reduce coupling. The breakout box itself, also acts as a source of RFI, presumably due to leakage RF noise from the EEG amplifier and was positioned as close as possible to the inside wall of the MSR. The minimization of the coupling between the hardware operating the SQUID system (including its data acquisition) and the stimulator unit requires physical separation. They are on opposite sides of the RF-shielded room. With these measures in place, concurrent single-channel MEG and 8-channel EEG following median nerve stimulation of a subject was conducted.

\begin{figure*}[!t]
    \centering
    \begin{minipage}[t]{\textwidth}
        \centering
        \includegraphics[width=\textwidth]{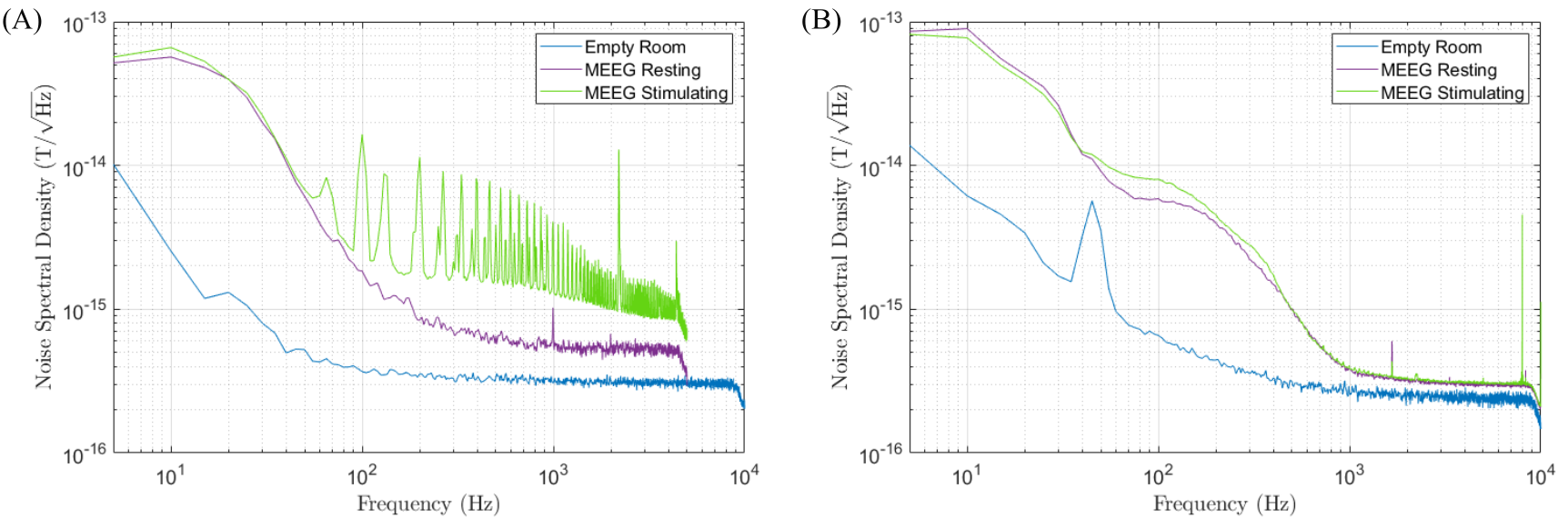}
        \caption{Noise spectral density of the MEG system for the (A) non- and (B) optimized setup. Note, for (A) the use of a second and then for (B) first order gradiometer leads to a white noise of approximately 300 aT/${\sqrt{\textrm{Hz}}}$ and 260 aT/${\sqrt{\textrm{Hz}}}$ in the empty room respectively. Operation of additional hardware (EEG and stimulation) leads to significantly worse noise performance before optimization of the setup minimizes this effect.}
        \label{fig:Figure3}
    \end{minipage}
\end{figure*}

Thanks to the described changes, a significant MEG noise reduction was achieved, thereby increasing the sensitivity of the device. The improvement can be seen when comparing A and B in Figure~\ref{fig:Figure3}. Here we show the noise spectral density (NSD) of the MEG system for various conditions. Though different subjects and gradiometer configurations were used between the recording sessions, one can clearly identify additional noise contributions. 

Resting state MEG is characterized, in addition to low frequency brain activity, by a higher white noise in the kHz range compared to an empty room measurement and stems from a combination of body noise and residual RF interference in the MSR~\cite{Storm2019}. For the non-optimized set-up, combination of MEG with simultaneous EEG results in significantly higher white noise which is caused by RFI of the non-optimized EEG set-up. During stimulation, the MEG noise increases dramatically. The white noise is now of the order of 1 fT/${\sqrt{\textrm{Hz}}}$, again caused by additional RFI. Discrete peaks at the power line harmonics are dominant. Note, this performance might be considered adequate for standard MEG systems. 

For the optimized set-up, in addition to the brain signals at low frequency, a slightly elevated white noise is observed with simultaneous operation of the EEG system during resting state. Upon stimulation, the white noise performance does not increase and remains at about 300 aT/${\sqrt{\textrm{Hz}}}$. For the resting state and the stimulation measurements, only discrete peaks of currently unknown origin at about 1.5~kHz and 8~kHz appear. However, the physiologically relevant $\sigma$-band is unaffected.

\subsection{Simultaneous bimodal single-trial detection of hfSERs}
\begin{figure}[h]
	\includegraphics[width=0.96\columnwidth]{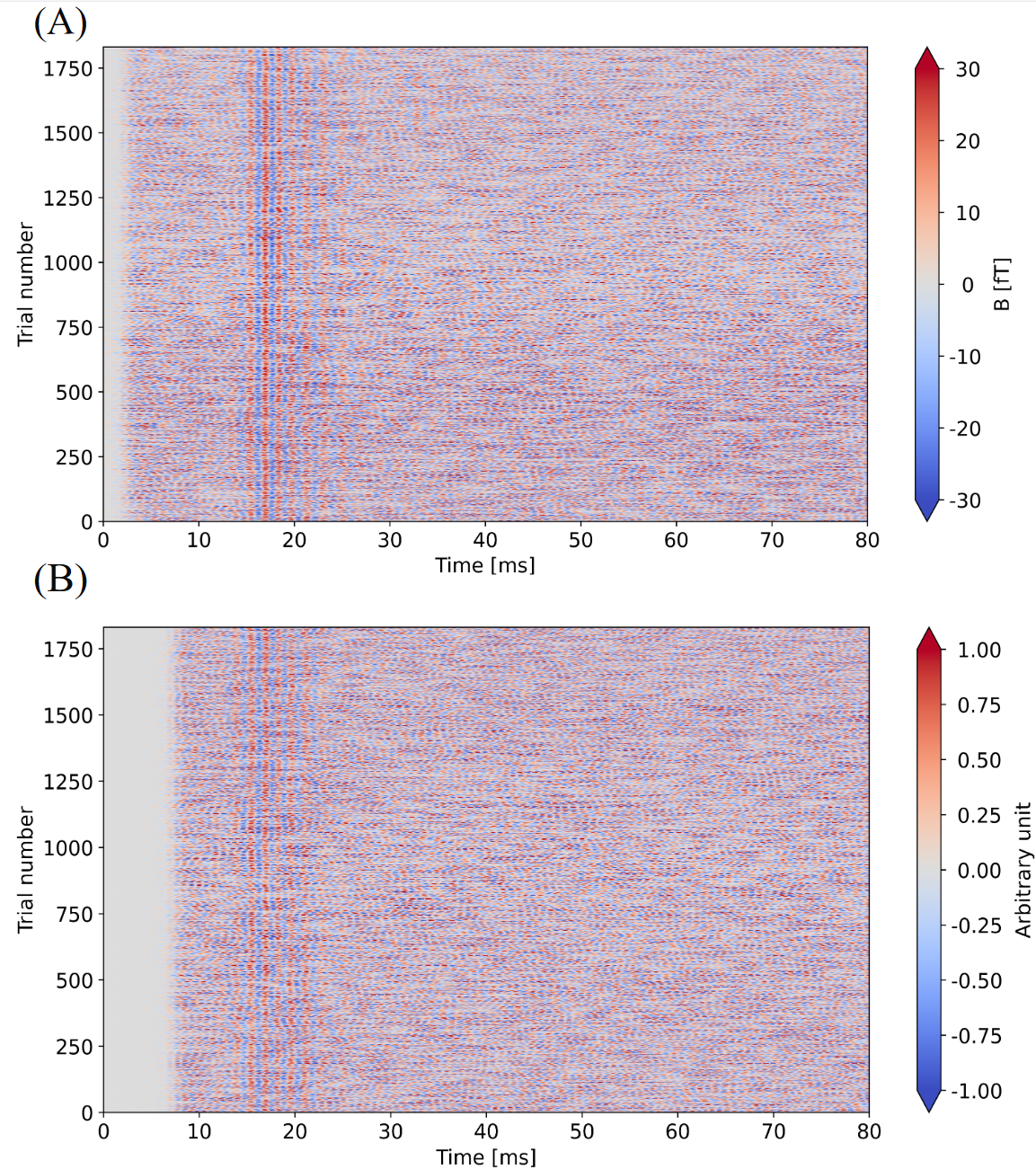}
	\caption{Time-amplitude resolved detection of single-trial hfSERs. Single trials after bandpass filtering (450 to 850 Hz) were
vertically stacked in chronological order with amplitudes coded in color saturation. (A) is MEG, (B) is EEG with the use of a CSP. Stimulation artifacts in the beginning of the single trials were digitally removed.}
	\label{img:Figure5}
\end{figure}

We can see the MEG/EEG results in this $\rm 450-850$~Hz range in Figure~\ref{img:Figure5}. Each stimulation epoch is stacked on top of each other, and these hfSERs can be seen at a single-trial level in both modalities. While the MEG data were single-channel, and show a maximum peak-to-peak value of approximately 50 fT, consistent with earlier work~\cite{Waterstraat2021}, the EEG data are a combination of all 8 channels. It uses a common spatial pattern algorithm (CSP) \cite{Ai2018} which maximizes discriminability between system noise and this $\sigma$ burst. The alignment of single-trial phases is clearly visible as formation of vertically oriented bands in the response window (15-30 ms). 

\section{Conclusion}
In this work, we demonstrated that careful hardware optimization allows mitigation of EMI impacting an ultra-low noise SQUID system which is caused by operating several electronic systems simultaneously with the highly sensitive device. Although the guidelines are particularly important for SQUID technology which is known to be susceptible to RFI, they can be applied to any sensitive magnetometry setup where ultra-low noise performance is required. The achieved improvements allowed us to measure hfSERs at the single-trial level in both MEG and EEG simultaneously. This work provides the technological foundation for our future bimodal ultra-sensitive multi-channel MEG/EEG instrumentation.

\textsf{\textbf{Author Statement}}\\
Research funding: Financial support by the German Research Foundation (DFG) is gratefully acknowledged (project 'SPIKE MEG', project no,: 511192033). Authors state no conflict of interest. Informed consent has been obtained from all individuals included in this study. This research has been given ethical approval (project 'SPIKE MEG' (PTB2022-2)).

\bibliographystyle{unsrt}
\bibliography{Litdata}

\end{document}